\documentclass[aoas,nameyear,dvips]{arximspdf}
\usepackage{graphics}

\doi{10.1214/09-AOAS34INTRO}
\volume{3}
\issue{4}
\pubyear{2009}
\firstpage{1233}
\lastpage{1235}

\begin{document}
\begin{frontmatter}

\title{Introducing the discussion paper\break by Sz\'{e}kely and Rizzo}
\runtitle{Introduction}
\pdftitle{Introducing the discussion paper by Szekely and Rizzo}

\begin{aug}
\author[A]{\fnms{Michael A.} \snm{Newton}\ead[label=e1]{aoas-man@biostat.wisc.edu}\corref{}}
\runauthor{M. A. Newton}
\affiliation{University of Wisconsin---Madison}
\address[A]{Department of Statistics\\
\quad and of Biostatistics\\
\quad and Medical Informatics\\
University of Wisconsin, Madison\\
1300 University Avenue\\
Madison, Wisconsin 53706\\
USA\\
\printead{e1}} 
\end{aug}




\end{frontmatter}

I recall a great sense of excitement in the seminar room in Madison
after Professor Sz\'{e}kely
presented the astonishing findings about distance covariance (in the spring of 2008).
It was one of the best statistics seminars I could remember.
Since before computers, statisticians have held up
Pearson's correlation coefficient as the most essential measure
of association between quantitative variables.  R. A. Fisher's reputation
was sealed, in part,  by solving the distribution of this statistic, and
so much of linear-model methodology relates to it.
 And all the time we've had to add the caveat about
independence following zero correlation \textit{only if}
the data are jointly normal.  Spearman's rank correlation has substantial
practical utility in cases where normality is unreliable, but the goal to have
a \textit{bona fide} dependence measure seemed to have been beyond the scope
of ordinary applied statistics.  Some valid measures did exist, but
being driven by empirical characteristic functions, they were too complicated
to enter the toolkit of the applied statistician.

Distance covariance not only provides a \textit{bona fide} dependence measure, but
it does so with a simplicity to satisfy Don Geman's \textit{elevator test} (i.e.,
a method must be sufficiently simple that it can be explained to a colleague
in the time it takes to go between floors on an elevator!).   You
 take all pairwise distances  between sample values of one  variable, and
do the same for the second variable. Then center the resulting distance
matrices (so each has column and row means equal to zero)
and average the entries of the matrix which holds componentwise products of
the two centered distance matrices.  That's the squared distance covariance between the two
variables.  The population quantity equals zero if and only if the variables
are independent, whatever be the underlying distributions and whatever
be the dimension of the two variables.  The depth of the
finding, the simplicity of the statistic, and the central role of statistical
dependence make this an important story for our discipline.

As a
numerical entree, consider six simulated examples of unusual joint
distributions, mimicking those
at the \href{http://wikipedia.org}{wikipedia.org} page
on Pearson correlation [R code is available in
supplementary files Newton (\citeyear{r2})].
In each case $n=500$ points are randomly sampled.  Although there is
dependence between horizontal and vertical components (in all but the case
on the far right), the Pearson correlation coefficient is essentially
zero (upper row),
consistently estimating the underlying zero correlation.  The
dependence is revealed by the distance correlation (lower row),
which is the normalized version of the distance covariance. As
expected, $p$-values
from the recommended Monte Carlo test of independence
 are all small,  except in the last case (not shown).

\begin{figure}

\includegraphics{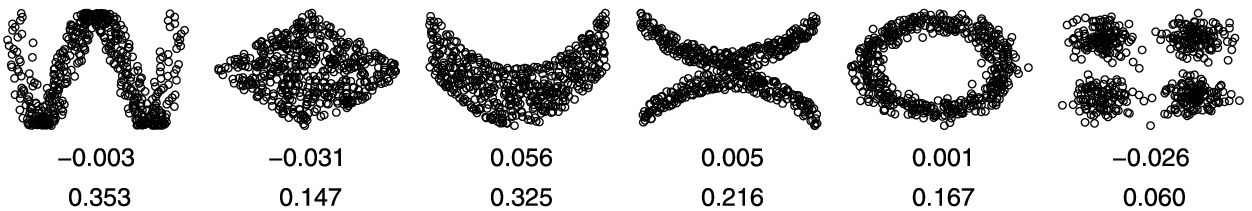}

\end{figure}
\begin{figure}[b]

\includegraphics{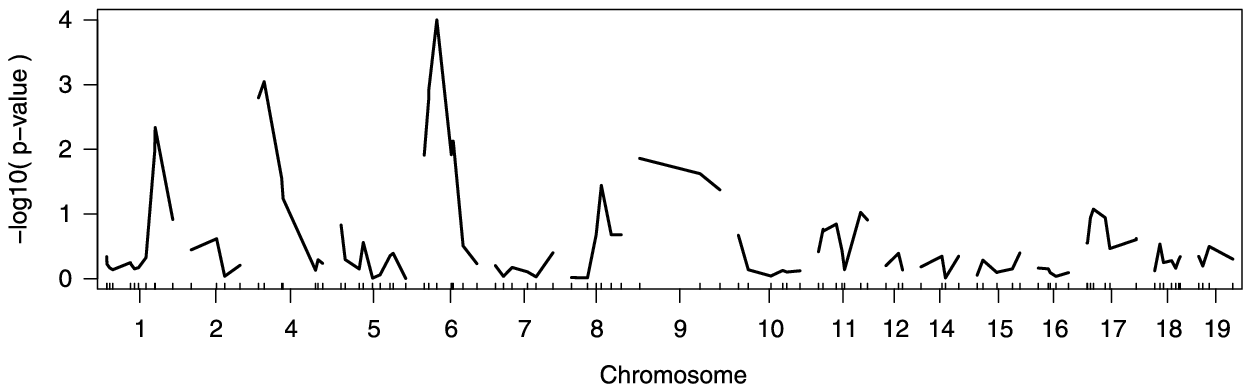}

\end{figure}

Work has just begun, I think, to explore the utility of distance covariance
in applied statistics.  In gene mapping, for example, the central statistical
problem is to identify dependencies between genetic information (genotype) and
other measured characteristics (phenotype) of sampled individuals.  Here
I describe one way that distance covariance could apply; many
versions seem possible.
Consider mapping a
 quantitative trait in the murine physiological response
to bacterial infection; 154~mice from a backcross population were typed
at 119 genetic markers in a study by Hopkins et al. (\citeyear{r1}).
A cell-based measure of response-to-infection
was also obtained in several tissues from the same animals.
At each genetic-marker position
 (horizontal axis), plotted in the figure below
is the $p$-value (negative log, base ten)
from the distance-covariance test of independence between the phenotype
(the infection response in bladder tissue)
and the genotype (a two-level covariate in this backcross population).
The distance between genotypes is the indicator
of distinct genotypes at one specific marker location, although extensions
could take advantage of various genome metrics.

In several earlier papers Professor Sz\'{e}kely and colleagues
introduced distance covariance and began to develop its theoretical
properties.
I invited them to prepare a paper for \textit{AOAS}, considering the
potential implications for applied statistics; the following work by
Professors Sz\'{e}kely and Rizzo is the response to this invitation.
It reports further properties
of the distance correlation based on a surprising
connection to Brownian motion, and
it presents some basic computational results from an R software implementation.
I am delighted that we have seven  contributions to a discussion of the paper which
explore the landscape of dependence in great detail.\looseness=1

\begin{supplement} [id-suppA]
\stitle{R code to simulate unusual joint distributions with zero Pearson correlation}
\slink[doi]{10.1214/09-AOAS34INTROSUPP}
\slink[url]{http://lib.stat.cmu.edu/aoas/34Intro/supplement.R}
\sdatatype{.R}
\sdescription{}
\end{supplement}

\def\bibname{Reference}

\printaddresses

\end{document}